\documentclass[aps,preprint,showpacs]{revtex4}
\usepackage{graphicx,epsfig}
\usepackage{amssymb}
\begin{document}

\title{Scattering and absorption sections of Schwarzschild--anti de Sitter with quintessence}

\author{Valeria Ram\'irez} 
\email{ra323273@uaeh.edu.mx}
\author{L. A. L\'opez}
\email{lalopez@uaeh.edu.mx}
\author{Omar Pedraza}
\email{omarp@uaeh.edu.mx}
\author{V. E. Ceron}
\email{vceron@uaeh.edu.mx}

\affiliation{ \'Area Acad\'emica de Matem\'aticas y F\'isica, UAEH, 
Carretera Pachuca-Tulancingo Km. 4.5, C P. 42184, Mineral de la Reforma, Hidalgo, M\'exico.}

\begin{abstract}
In this paper, we study the scattering and absorption sections of the Schwarzschild--anti de Sitter black hole surrounded by quintessence. The critical values of the cosmological constant and the normalization factor are obtained.  We describe the event horizons and the extremal condition of the black hole surrounded by quintessence. The effects of quintessence on the classical and semi--classical scattering cross--sections have been estimated. Also, the absorption section is studied with the sinc approximation in the eikonal limit. We consider the quintessence state parameter in the particular cases $\omega=-2/3$ and $\omega=-1/2$.
\end{abstract}

\maketitle

\section{Introduction}

The universe contains a high percentage of dark energy, and this energy is responsible for the accelerated expansion of the universe. The cosmological constant can play a very considerable role in the dynamics of the universe because it is the first and simplest explanation of dark energy \cite{Padmanabhan:2002cp}.

The evidence that the universe contains black holes (BHs) of different solar masses \cite{Begelman1898} has led to various studies and observations. In this sense, studying the consequences of the black holes coexistence with other types of matter or energy is essential. 

There are alternative models as candidates for dark energy, based on scalar fields as are quintessence \cite{Capozziello_2006}, phantom \cite{PhysRevLett.91.211301}, K--essence \cite{PhysRevLett.85.4438} among others.

Applying the ideas of Kiselev (2003) \cite{Kiselev:2002dx} that present a new static spherically symmetric exact solutions of the Einstein equations for quintessential matter surrounding a BH, different investigations have emerged. For example; the null geodesics for Schwarzschild  surrounded by quintessence in \cite{Fernando:2012ue} is addressed as well as Reissner--Nordstr\"{o}m surrounded by quintessence in \cite{Malakolkalami:2015tsa} also the thermodynamics of the black holes with quintessence are studied in \cite{Ghaderi:2016ttd} and Hayward black hole surrounded by quintessence in \cite{Pedraza:2020uuy}.

The null geodesics structure of massless particles of the Schwarzschild--anti de Sitter BH with quintessence is studied in \cite{Malakolkalami:2015cza}. Also, the dynamics of neutral and charged particles around the AdS Schwarzschild black hole surrounded by quintessence is discussed in \cite{Khan:2021nmv}, where authors applied Noether symmetries and the conservation laws to discuss the behavior of the geodesics, analyzing only the effective potential and the escape velocity. In both studies, the quintessence state parameter is considered only in a particular case.  

One the most useful and efficient ways to study the properties of BHs is by scattering or absorption of waves or test fields around them, recently the scattering and absorption sections of BHs surrounded by quintessence have been addressed in \cite{Lopez:2021ujg}. The physics of particle scattering from different kinds of BHs is an active topic because the cross--sections behavior with respect to the scattering angle is an important feature. Also, another important aspect of BHs is their accretion rate related to absorbing matter and fields. Accretion has an important role in the phenomenology of active galactic nuclei, also in the classical (highfrequency) limit, absorption cross--sections are directly related to the shadows of BHs.

For the mentioned above, in the present paper, we propose to study the scattering and absorption sections of  Schwarzschild--anti de Sitter BH surrounded by quintessence. The paper is organized as follows: In Sec. II  the Schwarzschild--anti de Sitter BH surrounded by quintessence is presented, the critical values of cosmological constant and normalization factor are shown. Also the event horizons and the extremal cases are analyzed. In section III, the expressions for the classical and semi--classical scattering sections are presented. IV we analyze the  scattering sections for $\omega=-2/3$ and $\omega=-1/2$ in detail. Section V takes the same values, the absorption section is obtained by sinc approximation. Finally, conclusions are given in the last section.

\section{Schwarzschild--anti de Sitter surrounded by quintessence}

When the cosmological constant is included in the Einstein equations, one obtains solutions that represent black holes with several asymptotic cases, among them, asymptotically anti--de Sitter BHs as  Schwarzschild--anti de Sitter. The spaces anti--de Sitter have recently attracted a lot of attention because the AdS/CFT correspondence \cite{Gubser:1998bc}, which is a correspondence between a gravitation theory and a quantum field theory, can be applied to strongly coupled systems. The timelike property of the AdS boundary is other characteristic interesting as compared to other asymptotic space--times.

Then using the ideas of  Kiselev \cite{Kiselev:2002dx} who proposed static and spherically symmetric solutions that describe BHs surrounded by quintessence. The solution of the Schwarzschild--anti de Sitter black hole surrounded by quintessence (Sch--aBH-$\omega$) is written as 

\begin{equation}\label{mfa}
ds^2=-f(r)dt^2+\frac{dr^2}{f(r)}+r^2d\theta^2+r^2\sin^2\theta d\phi^2\,,
\end{equation}
with

\begin{equation}\label{fSchw}
f(r)=1-\frac{2M}{r}-\frac{\Lambda r^{2}}{3}-\frac{C}{r^{3\omega+1}}\,.
\end{equation}

Where $M$ is the mass of the black hole and $\Lambda$ is the cosmological constant. The magnitude of $\omega$ which is the ratio of pressure to energy density takes values between  $-1<\omega<-1/3$.  The density of quintessence is always positive and given by:

\begin{equation}
\rho=-\frac{C}{2}\frac{3\omega}{r^{3(\omega+1)}}\,,
\end{equation}

where the constant $C$ is a positive normalization factor. For  $\Lambda=0$ we obtain the Schwarzschild black hole surrounded by quintessence that is studied in \cite{Fernando:2012ue}. 
  
The event horizons of the space--time (\ref{mfa}) are defined by the divergence of the metric function $g_{rr}$ that corresponds to the positive roots of the metric function $f(r)$ (\ref{fSchw}), for this analysis and the consequent analyses, we express  radial distance and  the parameter $C$ in units of mass as $r \to r/M$,  $C\to C/M^b$, in the case of the cosmological constant we consider $\Lambda \to \Lambda M^{2}$, with $b=3\omega+1$, then the event horizons are the roots of $3Cr + 6r^b - 3 r^{1+b} + r^{3+b} \Lambda =0$, the number of horizons depends entirely on the choice of the values of parameters $\omega$, $C$ and $\Lambda$.  

For the line element of Sch--aBH-$\omega$ (\ref{mfa}) to represent a black hole is necessary to determine the range of values of $\Lambda$ and $C$ such that $f(r)=0$. We use the method applied in \cite{PhysRevD.98.024015} \cite{Pedraza:2020uuy}, for determinate the ranges.

From the condition $f(r)=0$, we parametrize $\Lambda$ as a function of $r$ and $C$ as;

\begin{equation}\label{Lam}
\Lambda(r,C)= 3^{-(3+b)}(r^{1+b}-2r^{b}-Cr)\,.
\end{equation}

The cosmological constant has extrema $\left( \frac{d\Lambda(r,C)}{dr}=0\right) $ for a $C(r)$ given by;

\begin{equation}\label{C}
C(r)=\frac{2(r-3)r^{b-1}}{b+2}\,.
\end{equation}

$C(r)$ has a critical value (maximum) in certain $r_{crit}$, performing an analysis we obtain; 

\begin{equation}\label{rcr}
r_{crit}=\frac{3(b-1)}{b}\,.
\end{equation}

As $C>0$, then $C_{crit}(r_{crit})$ must be positive in the range $b\in(-2,0)$. So, the critical value of the quintessential parameter $C_{crit}=C(r_{crit})$ and the corresponding cosmological constant ($\Lambda (C_{crit},r_{crit})=\Lambda_{crit}$) are given by

\begin{equation}\label{Valcri}
C_{crit}=-\frac{2 (3 - 3/b)^b}{-2 + b + b^2},\quad \Lambda_{crit}=\frac{b^{3}}{18 - 27 b + 9 b^3}\,.
\end{equation}

In terms of $\omega$, Eq. (\ref{Valcri}) can be expressed as

\begin{equation}\label{Valcri-w}
C_{crit}=-\frac{2(9\omega)^{3\omega}}{(1+\omega)(1+3\omega)^{1+3\omega}},\quad \Lambda_{crit}=\frac{(1+3\omega)^{3}}{243\omega^{2}(1+\omega)}\,.
\end{equation}

\begin{figure}[ht]
\begin{center}
\includegraphics [width =0.45 \textwidth ]{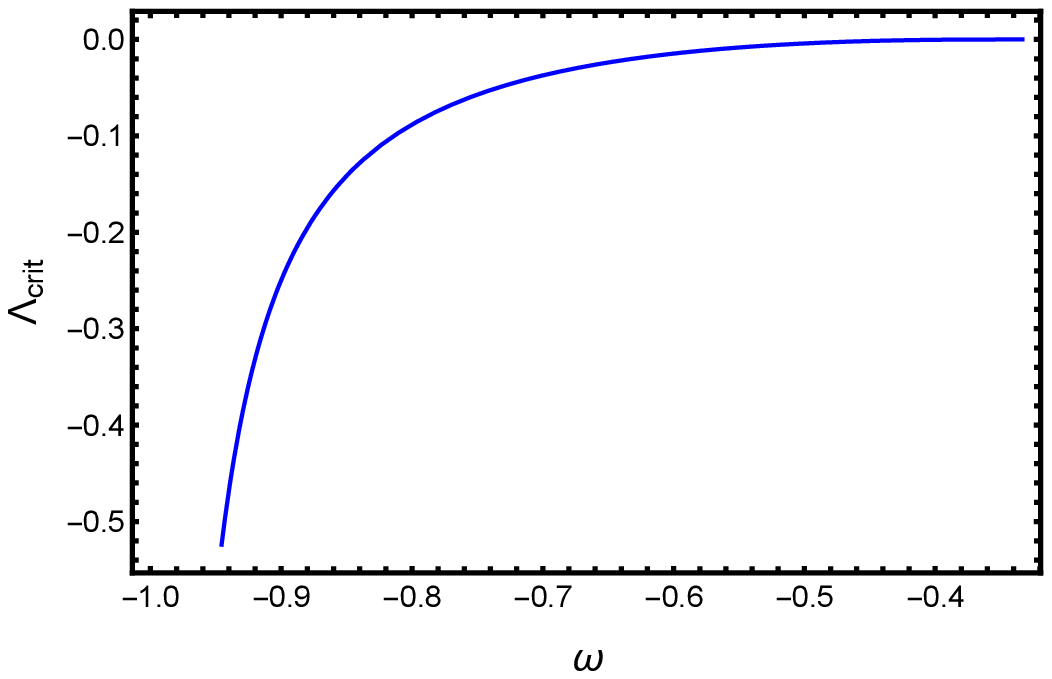}
\includegraphics [width =0.45 \textwidth ]{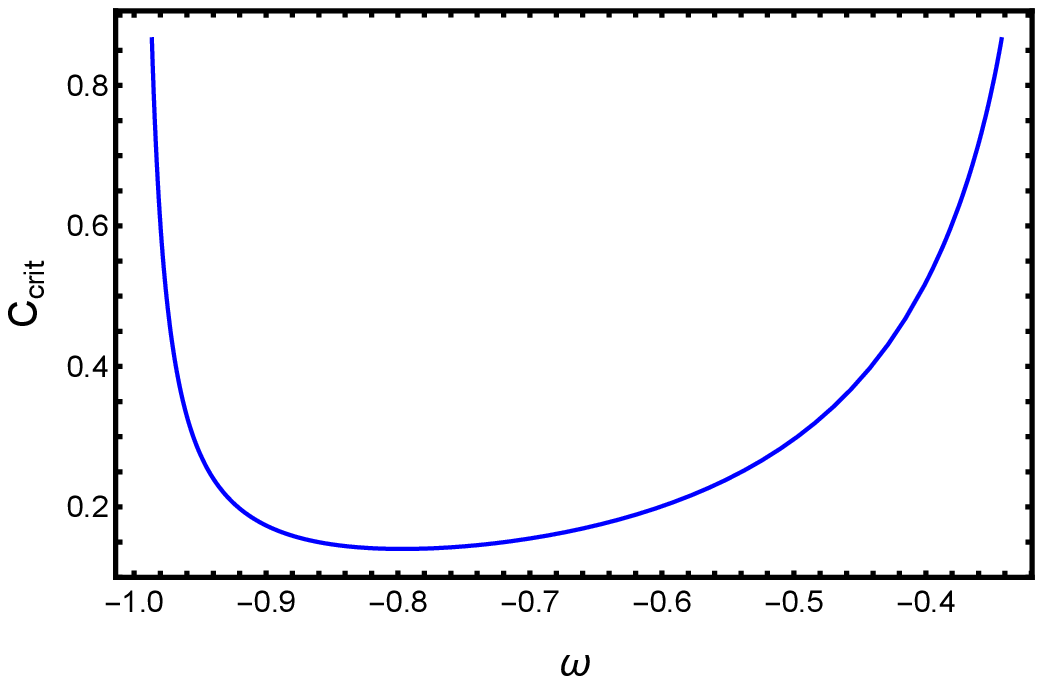}
\end{center}
\caption{The figures show the behavior of $\Lambda_{crit}$ and $C_{crit}$ in function of $\omega$ in the range  $-1<\omega<-1/3$. Notice that as $\omega$ grows, $\Lambda _{crit}$ also grows. $C_{crit}$ has a minimum for $\omega \approx  -0:796807$.}\label{Fig1}
\end{figure}

The behavior of the critical values $\Lambda_{crit}$ and $C_{crit}$ are shown in Fig. (\ref{Fig1}). The Sch--aBH--$\omega$ has horizons for values of $\Lambda_{crit} \leq \Lambda \leq 0 $ and $0 \leq C\leq C_{crit} $. The Fig. (\ref{Fig1}) shows that as $\omega$ increases, $C_{crit}$ decreases and presents a minimum (in $\omega \approx -0.796807$) then increases as $\omega$ increases.

On the other hand,when $\omega \rightarrow -1/3$,  we observe that $ \Lambda_{crit} \rightarrow 0$ (see Fig. \ref{Fig1}) means that a Schwarzschild BH surrounded by quintessence is formed.

In summary for  $\Lambda_{crit}\leq \Lambda \leq 0$ and $0 < C \leq C_{crit} $ the Sch--aBH-$\omega$ can represent a black hole with different horizons $r_{in}$, $r_{out}$ and $r_{\omega}$ (quintessence horizon). 

The extreme cases of Sch--aBH--$\omega$ can be obtained when the conditions $f(r)=0$ and $\frac{d}{dr}f(r)=0$ are satisfied simultaneously. Introducing the $\Lambda(r,C)$ of  (\ref{Lam}) in  $\frac{d}{dr}f(r)=0$, we obtained the condition $(2Cr+bCr+6r^b-2 r^{1+b})r^{-2-b}=0$ that can have two (or one) real roots denoted by $r_{+}$ and $r_{-}$.

In general, the extreme cases represent the situation when two or more horizons collapse, and the possible types (see \cite{PhysRevD.98.024015}) are:

\begin{itemize}

\item Type 1: The inner and outer horizons merge into a single horizon ($r_{in}=r_{out}$).

\item Type 2:  The outer horizon and quintessence horizon merge into a single horizon ($r_{out}=r_{\omega}$) .

\item Type 3:  The three horizons merge into a single horizon, and the Kiselev black holes is known as a super-extremal black hole.

\end{itemize}

We will focus on the particular case of $\omega -2/3$ the roots $r_{+}$ and $r_{-}$ are given by;

\begin{equation}
r_{\pm}=\frac{1 \pm \sqrt{1-6C}}{C}\,.
\end{equation} 

\begin{figure}[ht]
\begin{center}
\includegraphics [width =0.6 \textwidth ]{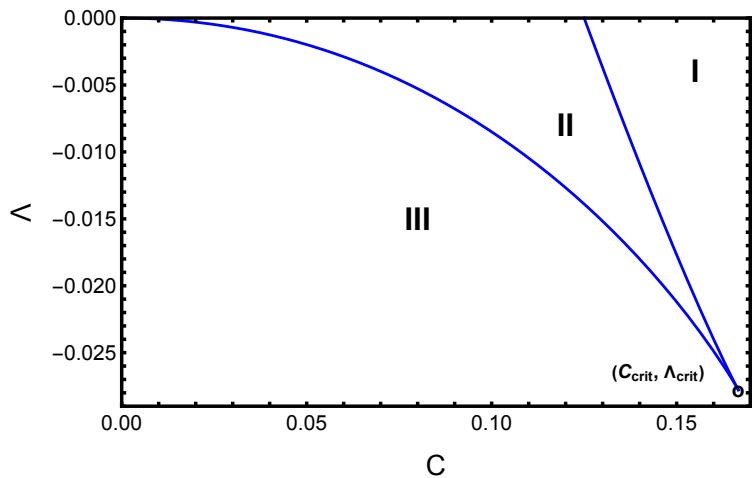}
\end{center}
\caption{The behavior of $\Lambda$  as a function of $C$ is shown for $\omega=-2/3$. The regions I, II and III contain the values of $\Lambda$ and $C$ so that Sch--aBH--$\omega$ has one horizon (regions I and III), three horizons (region II), and the boundary of the regions contains values that represent the extreme cases.}\label{Fig2}
\end{figure}

In Fig. (\ref{Fig2}) the behavior of $\Lambda$ as function of $C$ is shown for $\omega=-2/3$. In the regions I and III, the Sch--aBH--$\omega$ has one horizon, while in the region II there are three horizons. The boundary ($\Lambda(C,r_{-})$) of regions I and II as well as the boundary of regions II and III ($\Lambda(C,r_{+}$) represents the  extreme cases of of Sch--aBH--$\omega$ (Type 1 and 2). 

Depending on the values of the parameters $\Lambda$ and $C$ the number of the horizons may decrease from three to one, the cosmological horizon $r_{\omega}$ (quintessence horizon) never vanishes, then in this case  we say that the  Sch--aBH--$\omega$ describes a naked singularity.

For example, in \cite{Malakolkalami:2015cza} the null geodesics and the kinds of orbits are analyzed through the particular case $\omega=-2/3$ and Sch--aBH--$\omega$ has two horizons, the authors considered only the values of $C$ and $\Lambda$ located in the boundary of regions I and II. \cite {Khan:2021nmv} only considers a horizon for $\omega=-2/3$,  but do not specify ranges of the parameters where the horizon exists.

\begin{figure}[ht]
\begin{center}
\includegraphics [width =0.6 \textwidth ]{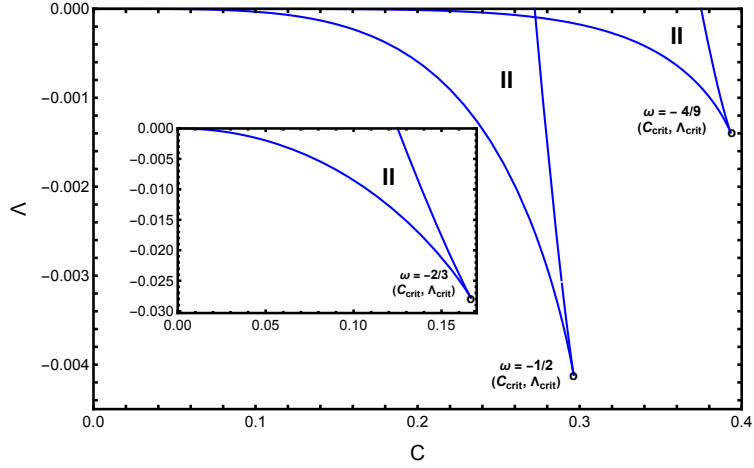}
\end{center}
\caption{The figure shows $\Lambda$ as a function of $C$  for different $\omega$; also, the critical values are shown. From the figure, we can see that for region II the values of $\Lambda$ for $\omega=-2/3$ are greater compared to $\omega=-4/9$.} \label{fig3}
\end{figure}

The development carried out for $\omega=-2/3$ can be applied for different values of $\omega$, however, the analysis must be numeric. The region II, where  Sch--aBH--$\omega$ has three horizons, is modified (see the Fig. (\ref{fig3}))  $II_{\omega=-2/3}>II_{\omega=-1/2}> II_{\omega=-4/9}$. 

The conditions for some values of $\omega$ are summarized in Table \ref{Tab1}, where the values of $\Lambda_{crit}$ and $C_{crit}$ are shown. It can be clearly shown how the factor $\omega$ modifies, the behavior of the horizons of Sch--aBH--$\omega$.

\begin{table}[!hbt]
\begin{center}
\begin{tabular}{|c|c|c|c|c|c|c|}
\hline 
$\omega$ & $-4/9$ & $-1/2$ & $-5/9$ & $-2/3$ & $-7/9$ & $-8/9$ \\ 
\hline 
$\Lambda_{crit}$ & $-1/720$ & $-1/243$ & $-2/225$ & $-1/36$ & $-32/441$ & $-125/576$ \\ 
\hline 
$C_{crit}$ & $0.3931$ & $0.2962$ & $0.2348$ & $0.1666$ & $0.140907$ & $0.16473$ \\ 
\hline 
\end{tabular} 
\caption{Critical values of the $\Lambda$ and $C$ for different $\omega$}\label{Tab1}
\end{center}
\end{table}

In all cases $\omega=-2/3$ , $\omega=-4/9$ and $\omega=-1/2$, the inner and outer horizons merge into a single horizon $r_{in}=r_{out}$, in boundary of regions I  and II. When  $r_{out}=r_{\omega}$  the boundary of regions II  and III are considered. Finally, if all horizons merge into a single horizon, and the Kiselev black holes is known as a super--extremal black hole.

\section{Classical and Semi--clasical Scattering cross sections}

The interaction of an incident plane wave with a black hole can be understood in terms of three quantities; the scattering cross section, absorption section, and polarization. 
	
Some gravitational scattering from astrophysical objects can be partially described in terms of the geometric optics through the analysis of geodesics. For example, the scatter of the light (null geodesic) that travels through space can be caused by a BH or other astrophysical objects. This phenomenon is known as gravitational lensing.

For the classical approximation of the  scattering cross--section, we consider that at very high frequencies, the incident wave propagates along the null geodesics \cite{Collins_1973}. 

The test particles that propagate along of null geodesics are described by the Lagrangian density $ \mathcal{L}=- \frac{1}{2}\dot{x}^{\mu}\dot{x}_{\mu}=0$, where "dot" denotes the derivative with respect to the affine parameter $\tau$. 

We restrict the geodesic motion to the plane $\theta=\pi / 2$ and solving for $\dot{r}^2$, we obtain $\dot{r}^2=E^2-V_{eff}$, where $V_{eff}=f(r)\frac{l^2}{r^2}$. The energy  $E=\frac{\partial \mathcal{L}}{\partial \dot{t}}$ and the angular momentum $l=-\frac{\partial \mathcal{L}}{\partial \dot{\phi}}$ of a test particle are conserved. As the impact parameter is defined as  $\hat b=l/E$, it is possible to obtain the impact parameter ($\hat b_{c}$) associated with the critical orbits, for any $\hat b>\hat b_c$ there is no scattering.

If we define $u=1/r$ and consider $\dot{r}^2=E^2-V_{eff}$, we can obtain the next expression;

\begin{equation}\label{Ang}
\left(\frac{du}{d\phi}\right)^{2}=\frac{1}{\hat b^{2}}-f(1/u)u^{2}\,.
\end{equation}

Differentiating (\ref{Ang}) with respect to $\phi$, we obtain;

\begin{equation}
\frac{d^{2}u}{d\phi^{2}}=-\frac{u^{2}}{2}\frac{df(1/u)}{du}-uf(1/u)\,.
\end{equation}

if we consider $\frac{d^{2}u}{d\phi^{2}}=0$ the positive root corresponds to the radius of the critical orbit for null geodesic $u_{c}$ and substituting in (\ref{Ang}), we obtain the critical impact parameter $\hat b_{c}$.

In the case of geodesics coming from infinity to a turning point $u_{0}$, the deflection angle is given by;

\begin{equation}\label{impact parameter}
\Theta\left(\hat b\right)=2\phi\left(\hat b\right)-\pi\,,
\end{equation}
where
\begin{equation}\label{Integral}
 \phi=\int_0^{u_0}du \left(\frac{1}{\hat b^2}-u^2 f(1/u)\right)^{-1/2}\,.
\end{equation}

The impact parameter $\hat b(\Theta)$, associated with the classical scattering cross--section is given by;

\begin{equation}\label{Sec}
 \frac{d\sigma}{d\Omega}=\frac{1}{\sin \Theta} \sum \hat b(\Theta) \left|\frac{d\hat b(\Theta)}{d \Theta}\right|\,.
 \end{equation}

For the scattering cross--section, we consider the null geodesics coming from infinity, and the incoming geodesics can orbit the BH several times before escaping.
 
Scattering waves from black holes produces diffraction effects that can be studied by analysis of geodesics. One of these effects is a glory which is a bright spot or halo that appears on--axis in the backward direction from the scatterer.
 
Now when we consider partial waves in the scattering phenomenon, it is necessary to consider the interference that occurs between partial waves with different angular momenta. This situation is not considered by the classical scattering cross--section (\ref{Sec}). The approximate method that considers the interference of the waves for low angles and high--frequency scalar plane waves ($\text{w} \gg 1$) is the semi--classical approach (Glory scattering) \cite{PhysRevD.31.1869}.  

The semiclassical approximation can be essential to determine the characteristics of the shadow of a black hole.

The glory approximation of the scalar scattering cross--section by spherically symmetric BHs is given by;

\begin{equation}
\frac{d\sigma_g}{d\Omega}=2\pi \text{w} \hat b_g^2 \left| \frac{d\hat b}{d\Theta}\right|_{\Theta=\pi}J_{2s}^2(\text{w} \hat b_g \sin\Theta)\,,
\end{equation}

with $\text{w}$ as the wave frequency. $J_{2s}^2$ stands for the Bessel function of first kind of order $2s$ where $s$ represents the spin, $s=0$ for scalar waves. The impact parameter of the reflected waves ($\theta \sim\pi$) is denoted by $\hat b_g$. As a semi--classical approximation, it is valid for $M\text{w} \gg 1$ ($M$ the mass of the BH).

\section{Classical and semi--classical scattering cross--sections for $\omega=-2/3$ and  $\omega=-1/2$ }

In this section, the classical and semi--classical scattering cross--sections are obtained for Sch--aBH-$\omega$. We analyze and compare the differences that may be similar but not identical. To carry out the analyzing scattering  cross--section, we consider $\omega=-2/3$ and  $\omega=-1/2$.

We propose to select the value of $\omega =-2/3$, which provides an intermediate range of  $C$ and $\Lambda$. The value $\omega =-1/2$ allows the Sch--aBH-$\omega$ to approach a Schwarzschild BH surrounded by quintessence. 
Also, the cases of $\omega =-1/2$  and $\omega =-2/3$,  enable a relatively simple treatment of the properties of Sch--aBH-$\omega$.

The $u_{c}$ (radius $r_{c}$) of the critical orbit for null geodesic is obtained of the equation (\ref{radi});

\begin{equation}\label{radi}
2-(2+b)Cu_{c}^{b}-6u_{c}=0\,,
\end{equation}

and the critical impact parameter $\hat b_{c}$ is given by;

\begin{equation}
M^{-2}\hat b_{c}^{2}=\frac{3}{3u_{c}^{2}-6u_{c}^{3}-3Cu_{c}^{2+b}- \Lambda}\,,
\end{equation}

we obtain the same $u_{c}$ that is reported in \cite{Malakolkalami:2015cza} for the case of $\omega=-2/3$ ($u_{c}=(1+\sqrt{1-6C})/6$), this result shows that Sch--aBH--$\omega$ with two and three horizons have the same circular orbit $r_{c}$ (photon sphere). Now for case $\omega=-1/2$, $u_{c}$ is given by;

\begin{equation}\label{r1/2}
u_{c}=\frac{27 A^{1/3} C^2}{-64 + 64 i \sqrt{3} + 16 A^{1/3} - A^{2/3} - i \sqrt{3} A^{2/3} + 
 648 C^2 - 648 i \sqrt{3} C^2}\,
\end{equation}

\begin{figure}[ht]
\begin{center}
\includegraphics [width =0.6 \textwidth ]{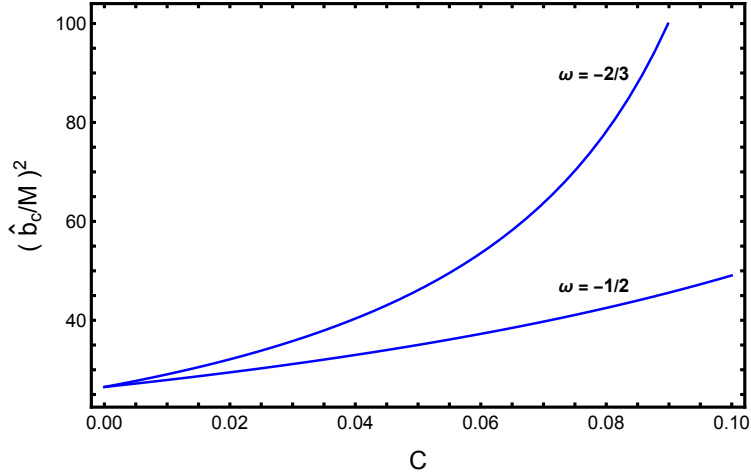}
\end{center}
\caption{The behavior of $(b_{c}/M)^{2}$ as function of $C$ is shown for different $\omega$ and $\Lambda=-0.002$, the
relative magnitudes of the critical impact parameters are $(\hat b_{c})_{-2/3}>(\hat b_{c})_{-1/2}$} \label{Fig3.1}
\end{figure}

Where $A=512-7776C^{2}+19683C^{4}+729\sqrt{729C^{8}-64C^{6}}$. It is possible to observe that the different $u_ {c}$ do not depend on the cosmological constant explicitly. Then the maximum of the effective potential is generated by $C$.
In the Fig. (\ref{Fig3.1}) the behavior of the critical impact parameter is shown for $\omega=-2/3$ and  $\omega=-1/2$, in both cases $\hat b_{c}$ increases as $C$ increases and in the limit $C\rightarrow 0$ both impact parameters tend to $27/(1-9\Lambda)$.

Some cases of $\omega$  allow  a relatively simple treatment of the properties of the geodesic equation of motion (\ref{Ang}), then in order to analyze the geodesic equation we consider $\omega=-2/3$, that can be written as;

\begin{equation}
\left(\frac{du}{d\phi}\right)^2=2\left(u-u_1\right)\left(u-u_2\right)\left(u-u_3\right)\,.
\end{equation}

Where $u_{i}$ are the roots  of (\ref{Ang}) given by;

\begin{eqnarray}
u_1&=&\frac{\left(1-\sqrt{3}i\right)\left(\frac{3C}{2}-\frac{1}{4 }\right)}{2^{3/2}3\Delta}
-\frac{\left(1+\sqrt{3}i\right)\Delta}{ 2^{1/3}6}+\frac{1}{6}\,,\\
u_2&=&\frac{\left(1+\sqrt{3}i\right)\left(\frac{3 C}{2 }-\frac{1}{4 }\right)}{2^{3/2}3\Delta}
-\frac{\left(1-\sqrt{3}i\right)\Delta}{ 2^{1/3}6}+\frac{1}{6 }\,,\\
u_3&=&-\frac{2^{1/3}\left(\frac{3 C}{2}-\frac{1}{4}\right)}{3\Delta}
+\frac{\Delta}{ 2^{1/3}3}+\frac{1}{6}\,,\\
\Delta&=&\left(\sqrt{\delta+4 \left(\frac{3 C}{2}-\frac{1}{4}\right)^3}-\frac{27}{2 \hat b^2 M^{-2}}-\frac{9 C}{4 }+\frac{1}{4}-\frac{9 \Lambda }{2 }\right)^{1/3}\,,\\
\delta&=&\left(-\frac{27}{2 b^2 }-\frac{9 C}{4}+\frac{1}{4 }-\frac{9 \Lambda }{2 }\right)^2\,,
\end{eqnarray}

and $u_{1}<u_{2}<u_{3}$. Then (\ref{Integral}) take the form;

\begin{equation}
\int_0^{u}\frac{du}{\sqrt{2\left(u-u_1\right)\left(u-u_2\right)\left(u-u_3\right)}}=-\frac{2}{\sqrt{2\left(u_2-u_1\right)}}F\left(\xi,y\right)\,,
\end{equation}

with, the incomplete elliptic integral

\begin{equation}
F\left(\xi,y\right)=\int_0^{\sin\xi}\frac{dx}{\sqrt{1-x^2}\sqrt{1-x^2y^2}}\,,
\end{equation}

where,
\begin{equation}
\xi=\arcsin\sqrt{\frac{u_2-u_1}{u-u_1}},\quad 
y=\sqrt{\frac{u_1-u_3}{u_1-u_2}}\,.
\end{equation}

Finally we obtain.

\begin{equation}\label{angel}
\phi=-\frac{2}{\sqrt{2M\left(u_2-u_1\right)}}F\left(\xi,y\right)+\phi_0.
\end{equation}

Solving (\ref{angel}) numerically with appropriate boundary conditions is possible to obtain the null geodesic followed by Sch--aBH--$\omega$. In the case of $\omega=-1/2$ the analytic computation of (\ref{angel}) is almost impossible, then the solution is numerically obtained.

\begin{figure}[ht]
\begin{center}
\includegraphics [width =0.4 \textwidth ]{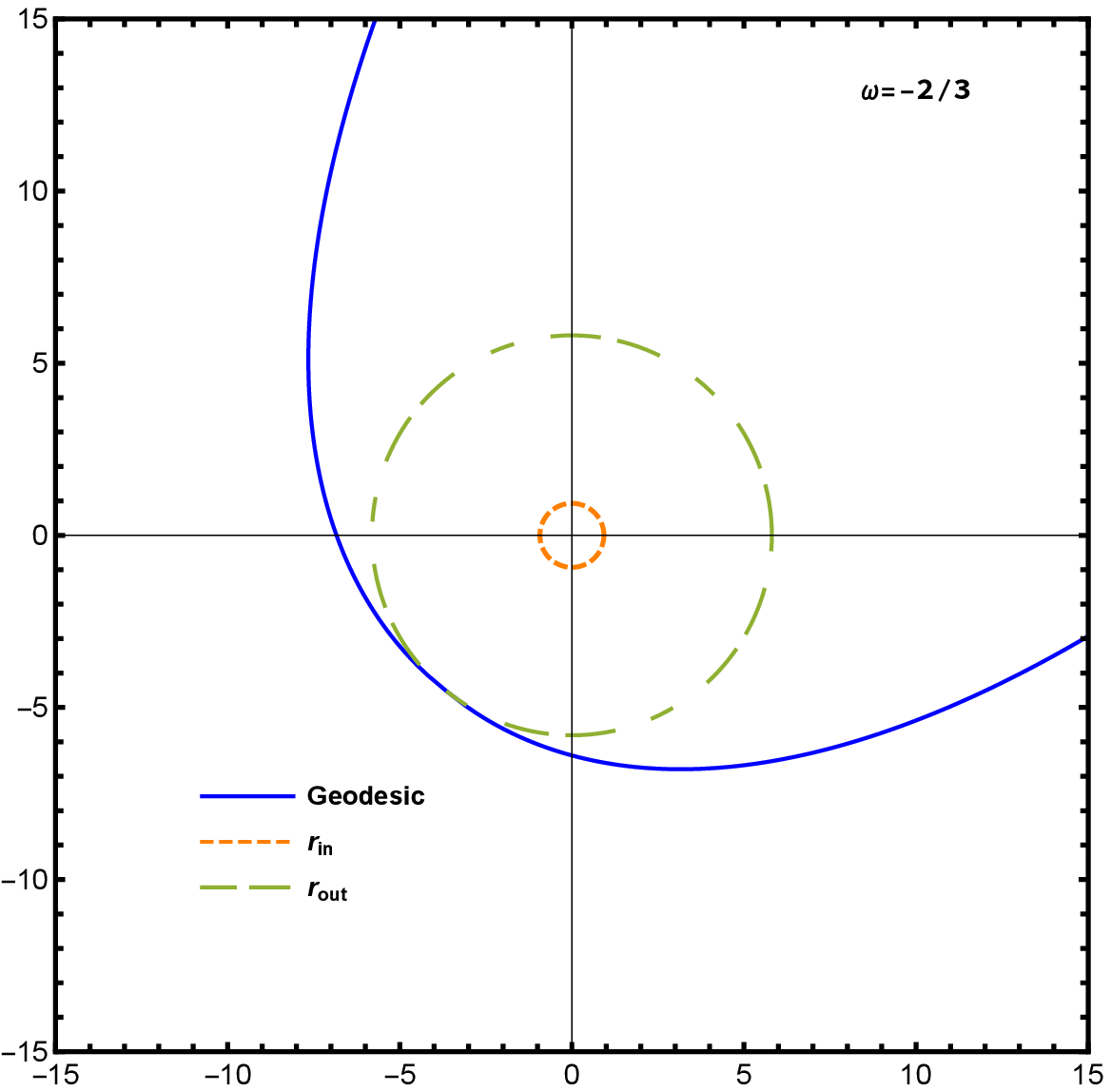}
\includegraphics [width =0.39 \textwidth ]{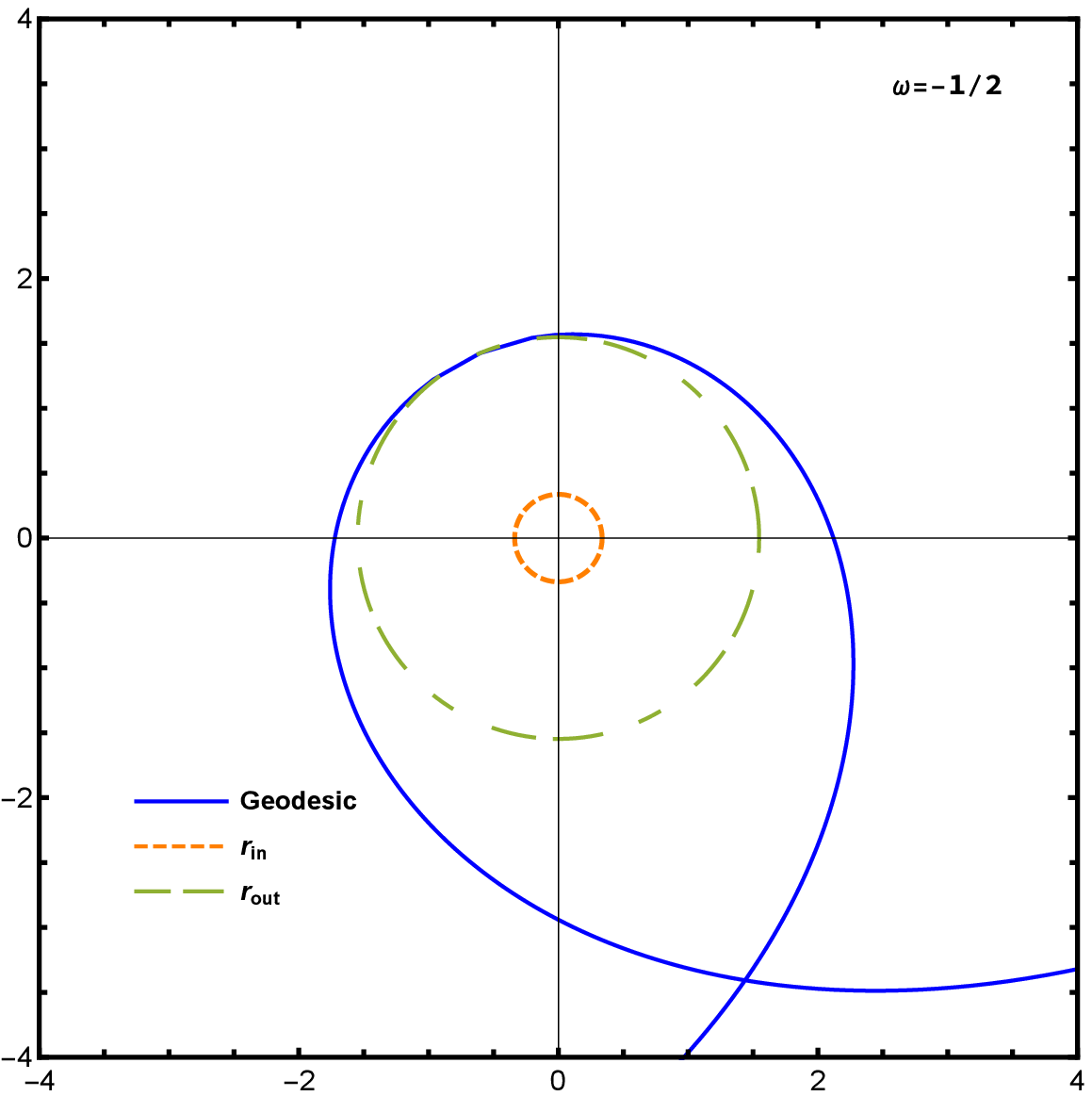}
\end{center}
\caption{The figures show the behavior of geodesics approaching a Sch--aBH--$\omega$ from infinity, for $\omega=-2/3$ and $\omega=-1/2$.}\label{Fig3.2}
\end{figure}

The corresponding motion of geodesics is given in the Fig. (\ref{Fig3.2}), and shows the scattering of geodesics, as  $r_{\omega}>r_{out}$ then the movement of the null geodesics can be located within the quintessence horizon (apparent horizon), but if it passes within the horizon $r = r_{out}$, the geodesics could fall into the BH. The above is considered to obtain the classical and semi--classical cross--sections.

To analyze the scattering of the null geodesics, we consider that the effective potential has asymptotic value $E_{min} \approx l^2\Lambda/3$ when $r\to\infty$ and a maximum located at $r_{c}$, then the energy values of test particles are considered between these values.

\begin{figure}[ht]
\begin{center}
\includegraphics [width =0.45 \textwidth ]{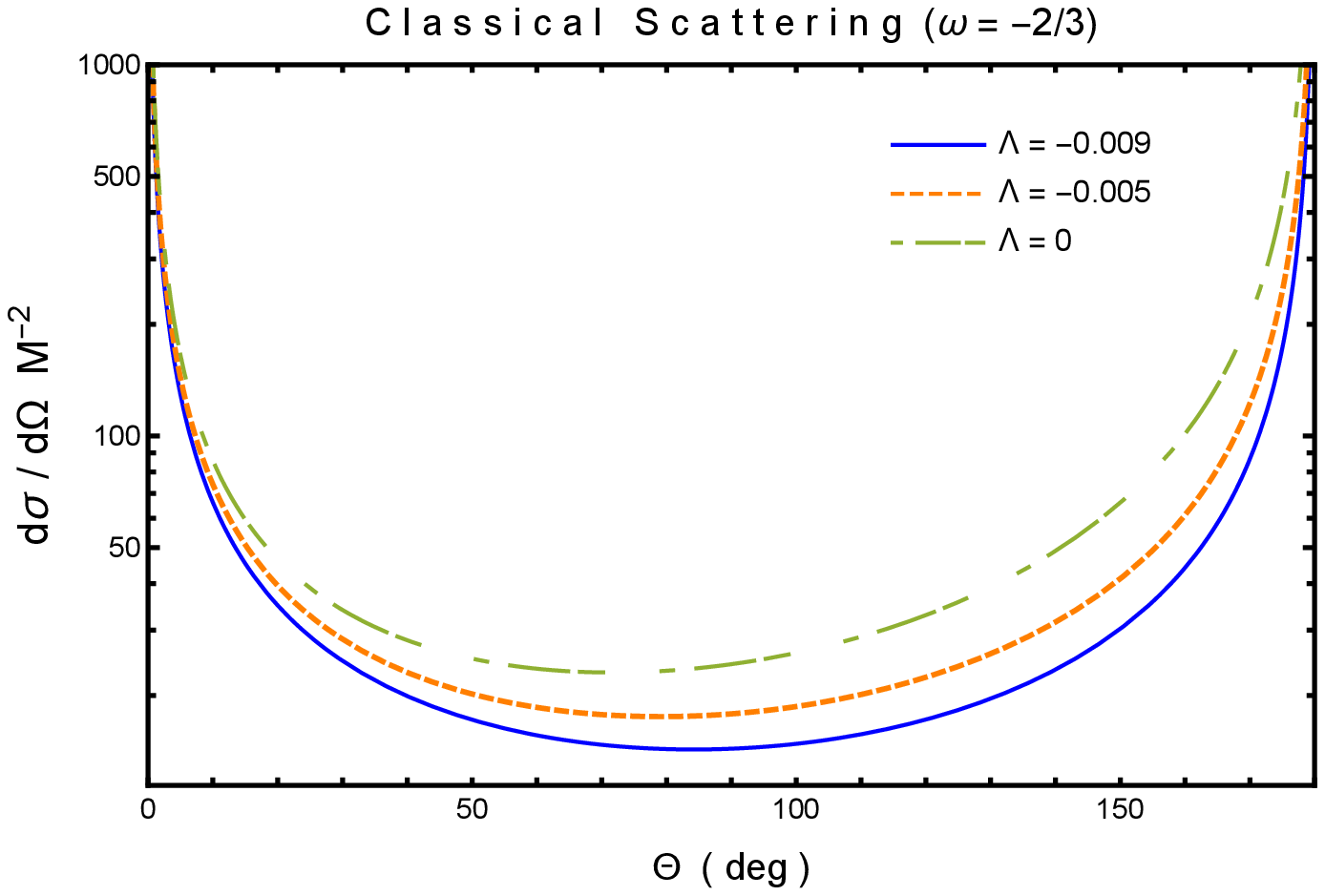}
\includegraphics [width =0.45 \textwidth ]{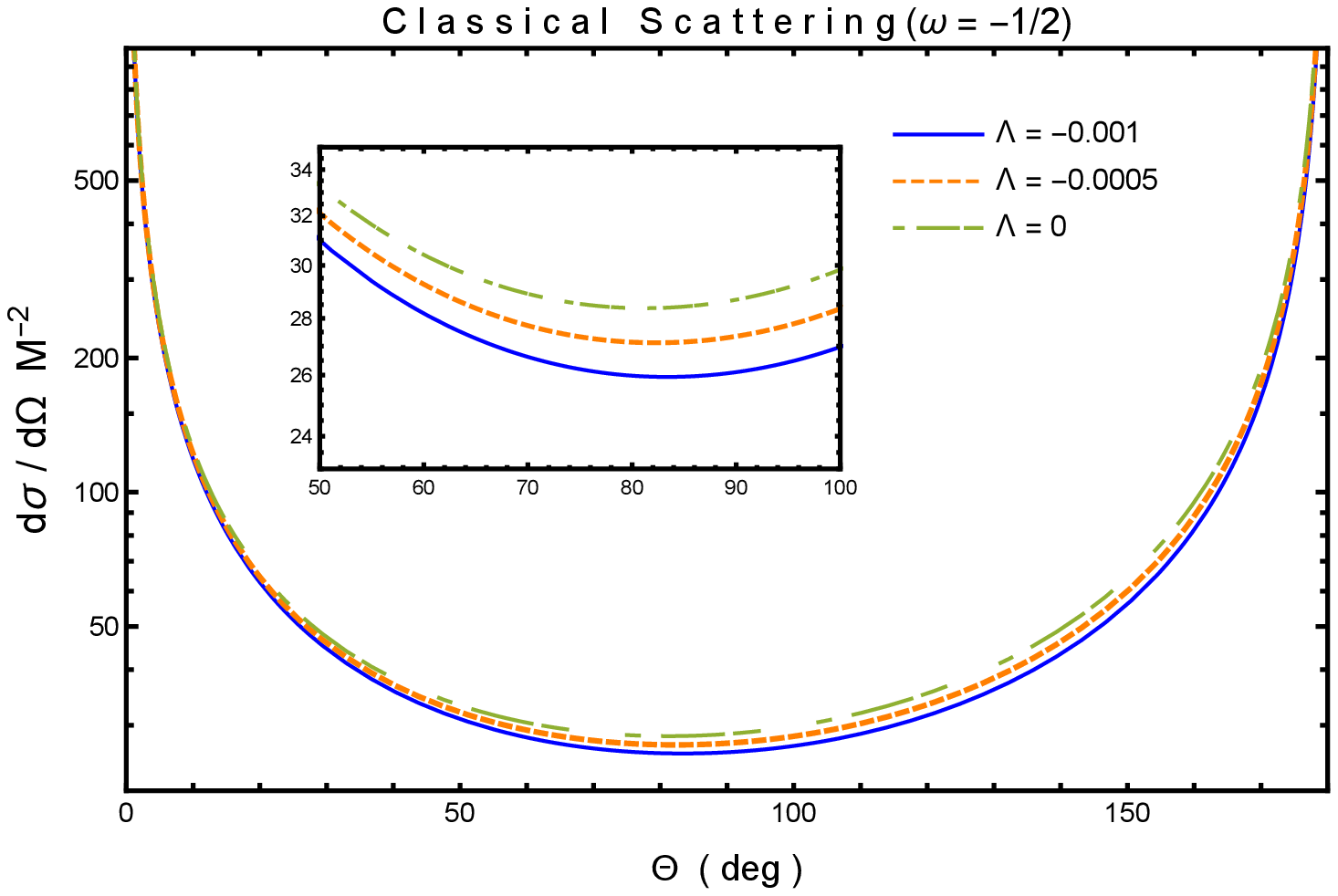}
\end{center}
\caption{The behavior of classical scattering cross--section is shown varying the value of $\Lambda$, in the first panel, we plot the cross--section for $\omega=-2/3$ with $C=0.12$ and in the second panel, the cross-section is plotted for $\omega=-1/2$ and $C=0.25$ in the small box the difference between the cross--sections for $\omega=-1/2$ are shown.}\label{Fig4}
\end{figure}

In Fig (\ref{Fig4}), the classical cross--sections for Sch--aBH--$\omega$ are compared considering different values of $\Lambda$ and $C$ in the case that Sch--aBH--$\omega$ has three horizons. It is possible to note that in both  cases ($\omega=-2/3$ and  $\omega=-1/2$) there is no significant difference between in the case of the classical scattering cross--section for small and large angles. 

Comparing the scattering cross--sections from Sch--aBH--$\omega$ and Schwarzschild BH surrounded by quintessence ($\Lambda=0$), we observe that  the cross--section of  Schwarzschild BH surrounded by quintessence is greater than Sch--aBH--$\omega$, for both cases $\omega=-2/3$ and $\omega=-1/2$.

\begin{figure}[ht]
\begin{center}
\includegraphics [width =0.45 \textwidth ]{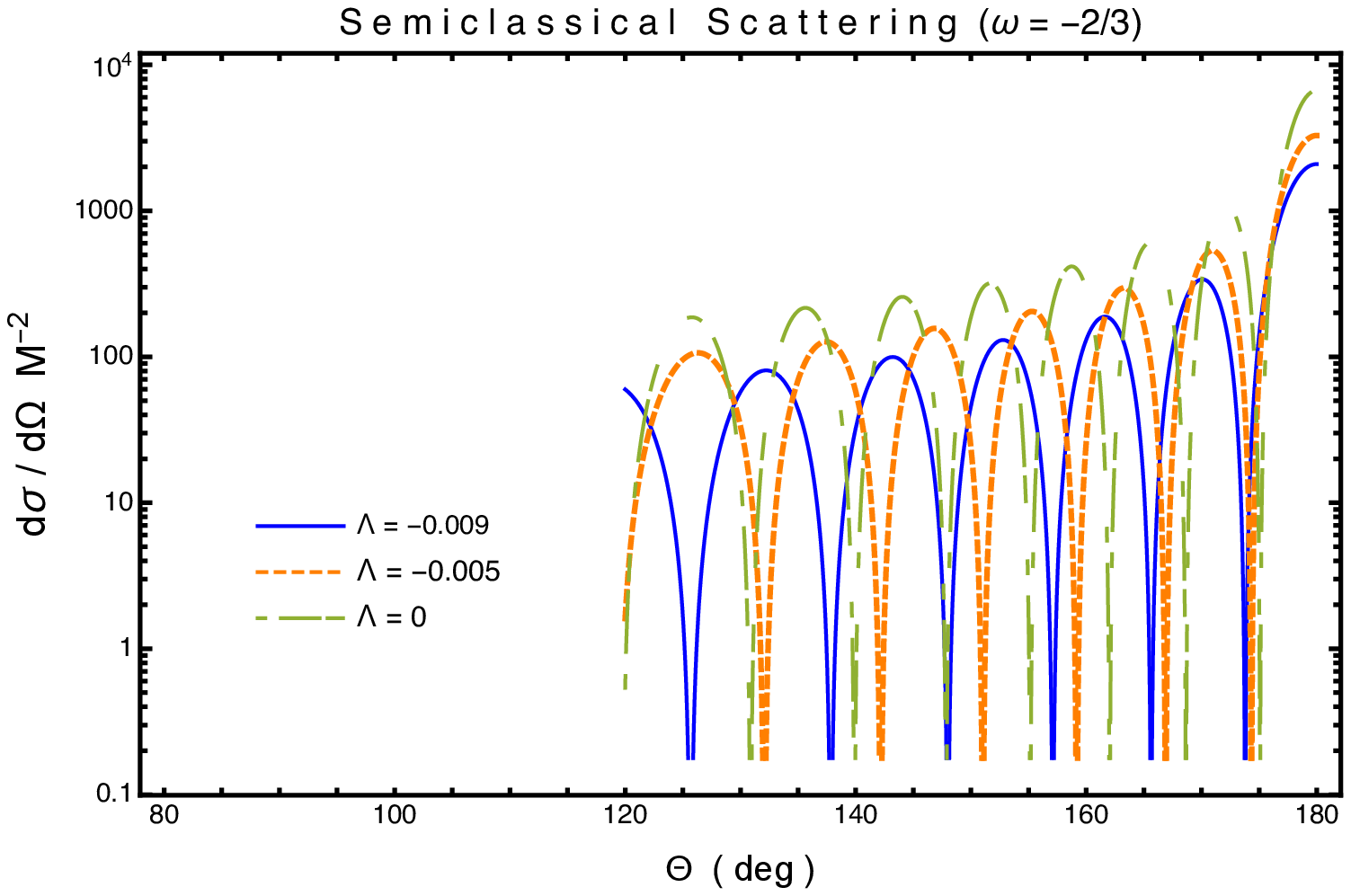}
\includegraphics [width =0.45 \textwidth ]{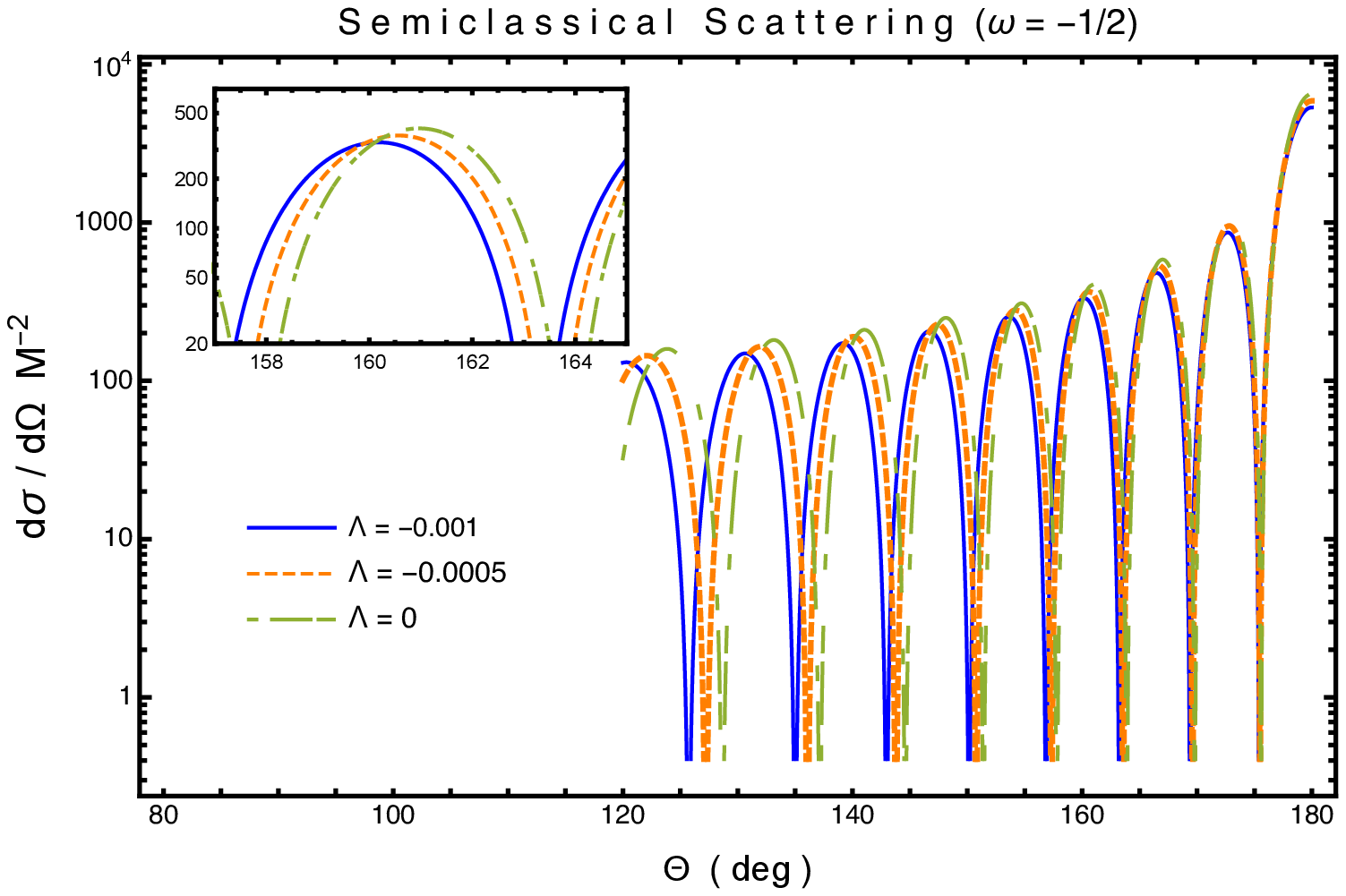}
\end{center}
\caption{The behavior of semi--classical scattering cross--section is shown varying the value of  the cosmological constant $\Lambda$, in the first panel, we plot the for $\omega=-2/3$ with $C=0.12$ and the second panel, the semi--classical cross--section is plotted for $\omega=-1/2$ and $C=0.25$ in the small box the difference between the semi--classical cross--sections for $\omega=-1/2$ are shown, in both cases we consider $\text{w} M =2$.}\label{Fig5}
\end{figure}

The  semi--classical scattering differential cross--sections for Sch--aBH--$\omega$ with $\omega=-2/3$ and $\omega=-1/2$ are shown in the Fig. (\ref{Fig5}). It can be observed that in the case of Sch--aBH--$\omega$ the scattering section is highly similar to Schwarzschild BH surrounded by quintessence ($\Lambda=0$) section. 

In the semi-classical results, we see that the interference fringes widths increase with $\omega=-2/3$ and in the case of $\omega=-1/2$  decrease, but the difference between Sch--aBH--$\omega$ and Schwarzschild with quintessence is more notable with $\omega=-2/3$.

Regarding the cross--sections, it is possible to mention that when we vary the normalization factor $C$, the behavior of cross--sections is similar if we fix $C$ and vary $\Lambda$. It is also possible to mention that in general in the classical and semi--classical cross--section $(\frac{d\sigma}{d\Omega})_{\Lambda = 0} > (\frac{d\sigma}{d\Omega})_{\Lambda \neq 0}$ i.e  that $\Lambda$ attenuates the effect of the quintessence in the scattering sections.

\section{Absorption cross--section}

In the approximation of low frequencies, the absorption cross--section by spherically symmetric black holes equals the BHs horizon area \cite{Higuchi:2001si}. In the approximation of high frequencies, the absorption cross--section can be considered as the classical capture cross--section of the null geodesics in massless scalar waves. In this limit the absorption cross section is also called geometric cross--section $\sigma_{geo}=\pi\hat b_{c}^{2}$.

On the other side, it is well known that the absorption cross--section oscillates around the limiting value of the geometric cross-section. This fluctuation is a generic feature of black holes endowed with a photon sphere, which can be described in terms of the area of the geometrical cross--section and the properties of the waves trapped near the photon sphere.

In \cite{PhysRevD.83.044032} the author showed that  in the case of  high frequencies (eikonal limit) the oscillatory part of the absorption cross--section can be written in terms of the parameters of the unstable null circular orbits. 

The oscillatory part of the absorption cross--section in the eikonal limit is defined by; 

\begin{equation}
\sigma_{osc}= - 4\pi \frac{\lambda\hat b_{c}^{2}}{\text{w}}e^{-\pi \lambda\hat b_{c}}\sin\left(\frac{2\pi \text{w}}{\Omega_{c}}\right)\,,
\end{equation}

where $\lambda$ is the Lyapunov \cite{Cardoso:2008bp} exponent given by;

\begin{equation}\label{Lyapunov}
\lambda^{2}=\frac{f(r_{c})}{2r_{c}^{2}}\left[2f(r_{c})-r_{c}^{2}f^{''}(r_{c})\right]\,,
\end{equation}

$r_{c}$, the radius of the unstable null circular orbit and $b_c$ is the critical impact parameter.

while the orbital angular velocity  is ;

\begin{equation}\label{angular1}
\Omega_{c}= \sqrt{\frac{f_c}{r_c^2}}\,.
\end{equation}

Then the absorption cross--section in the limit of high frequencies is proportional to the sum of $\sigma_{osc}$ and $\sigma_{geo}$ (the sinc approximation)

\begin{equation}
\sigma_{sinc} \approx \sigma_{geo} + \sigma_{osc}\,.
\end{equation}

\subsection{Absorption cross--section with $\omega=-2/3$ and  $\omega=-1/2$ }

In this subsection, the absorption cross-sections in the sinc approximation are obtained for Sch--aBH--$\omega$. We analyze and compare the differences that may be similar but not identical. 

The Lyapunov exponent for $\omega=-2/3$ is given by;

\begin{equation}
M^{2}\lambda^{2}_{\omega=-2/3}=\frac{(Cr_{c}-1)(6-3r_{c}+3Cr_{c}^{2}+r_{c}^{3}\Lambda)}{3r_{c}^{3}}\,,
\end{equation}

with $r_{c}$ is reported in \cite{Malakolkalami:2015cza} as previously mentioned. The Lyapunov exponent for $\omega=-1/2$ is given by;

\begin{equation}
M^{2}\lambda^{2}_{\omega=-1/2}=\frac{(9C\sqrt{r_{c}}-8)(6-3r_{c}+3Cr_{c}^{3/2}+r_{c}^{3}\Lambda)}{24r_{c}^{3}}\,,
\end{equation}

the $r_{c}$ for the case $\omega=-1/2$ were obtained in (\ref{r1/2})

\begin{figure}[ht]
\begin{center}
\includegraphics [width =0.45 \textwidth ]{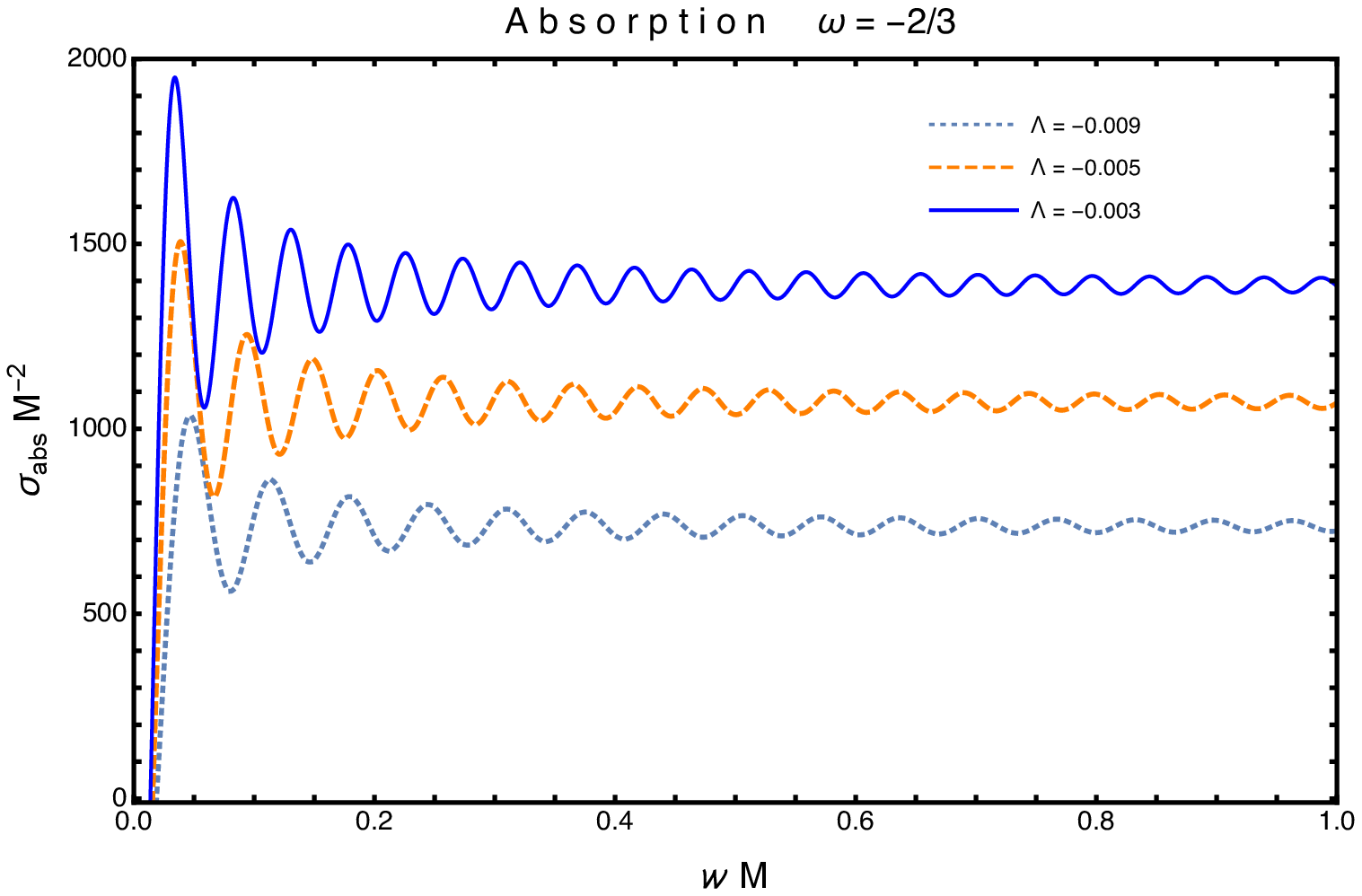}
\includegraphics [width =0.45 \textwidth ]{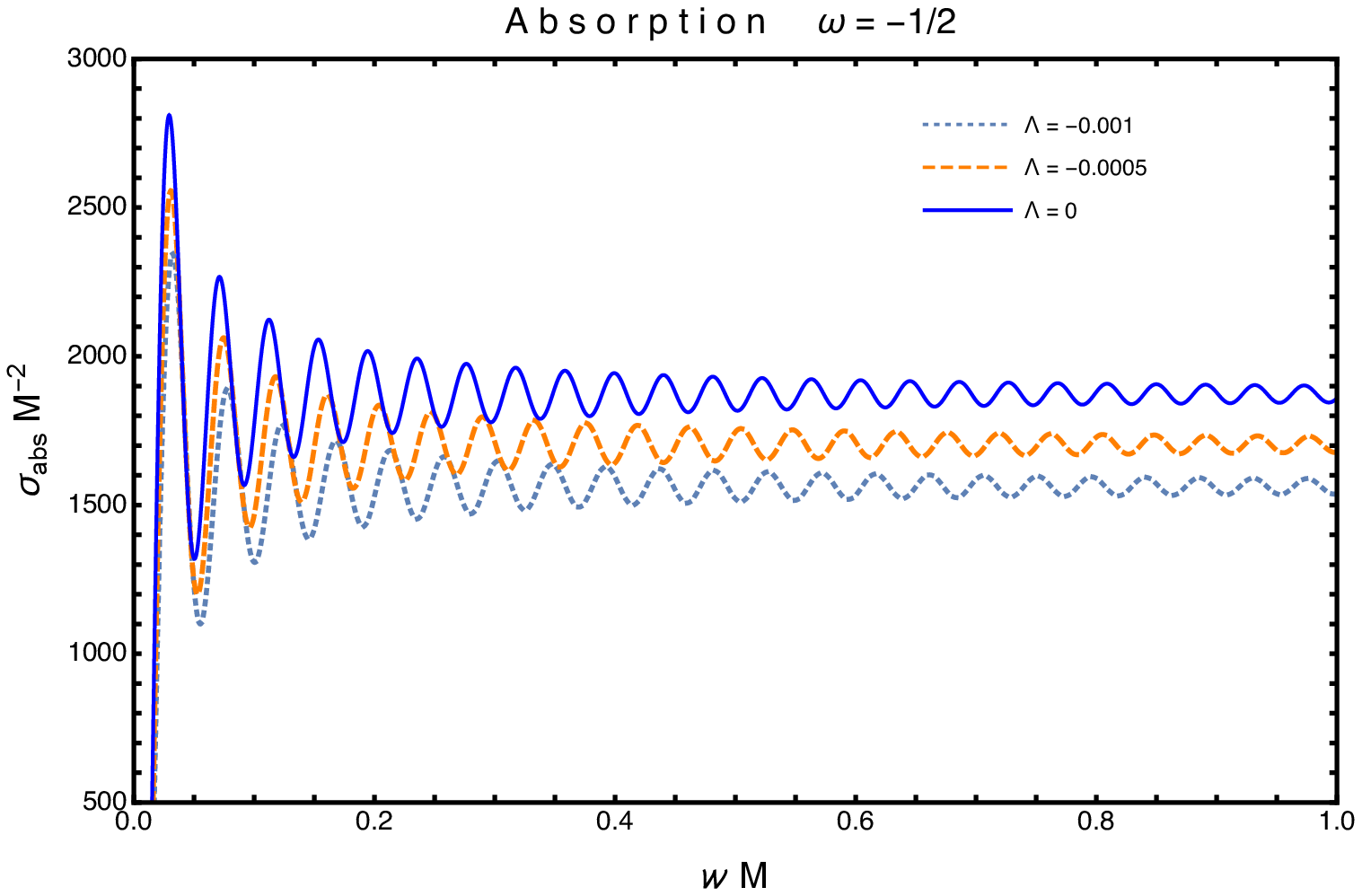}
\end{center}
\caption{The behavior of absorption cross section in the sinc limit is shown varying the value of the cosmological constant $\Lambda$, in the first panel, we plot the absorption cross section for $\omega=-2/3$ with $C=0.12$ and the second panel, the absorption cross--section is plotted for $\omega=-1/2$ and $C=0.25$}\label{Fig6}
\end{figure}

The absorption cross--sections for $\omega =-2/3$ and $\omega =-1/2$ are plotted in Fig. (\ref{Fig6}), respectively in the case of the sinc approximation. It is obvious that the amplitude of absorption cross--section tends to zero as $w M$ increases. It also shows that the difference of the curves is numerically significant  with respect to their amplitude with larger values of $w M$ in both cases.

Moreover, for each value $\omega$, it is possible to mention that $\sigma_{-1/2}>\sigma_{-2/3}$, also  the corresponding absorption cross--section starts from zero, reaches a maximum value $\sigma_{Abs}$, and decreases asymptotically. Finally,  when we consider $\omega=-1/2$, the behavior shows that differences occur on a smaller scale than with $\omega=-2/3$.

\section{Conclusions}

In this paper, we show how the scattering and absorption sections are modified by considering the effects quintessence in the Schwarzschild--anti de Sitter black hole.

First, we present the critical values of the $C_{crit}$ (normalization factor) and cosmological constant ($\Lambda$ )  in terms of $\omega$, when  $\Lambda_{crit}\leq \Lambda <0 $ and $ 0  \leq C \leq C_{crit} $ the solution of Schwarzschild--anti de Sitter surrounded by quintessence (Sch--aBH--$\omega$) has one, two, or three horizons. The regions where the Sch--aBH--$\omega$ has one, two, or three horizons are analyzed in the particular cases $\omega=-2/3, -1/2, -4/9 $ and we see that the behavior is similar for all values. Also the extreme cases of Sch--aBH--$\omega$ are  studied. When $\omega \rightarrow -1/3$ we observe that a Schwarzschild BH surrounded by quintessence is formed and  $C_{crit}$ presents a minimum in $\omega \approx -0.796807$.

We fixed  $\omega=-2/3$, which provides an intermediate range of  ($C$, $\Lambda$) and $\omega=-1/2$ allowing the Sch--aBH-$\omega$ to be approaches as a Schwarzschild BH surrounded by quintessence (Sch--BH--$\omega$), we have also focused our attention in these cases that enable a relatively simple treatment of the properties. For the analysis of cross--sections we consider the ranges of the cosmological constant and the normalization factor where the Sch--aBH--$\omega$ has three horizons. We observe that the classical and semi--classical scattering cross--sections for Sch--BH--$\omega$ are greater than the ones for Sch--aBH--$\omega$. We may conclude that in both cases, the cosmological constant attenuates the effect of the quintessence in the scattering sections. Also, for the classical scattering cross--section the most important differences come from big angles. For the semi--classical scattering cross--section, we observed that the interference fringes widths increase in the case of $\omega=-2/3$. Then, the quintessence state parameter is the cause of high interference. For $\omega=-2/3$, the differences between the cross--sections are more noticeable since the impact parameters are more significant than the ones for $\omega=-1/2$ . 

For the absorption cross--section the sinc approximation is used, so that the absorption section is given in function of the Lyapunov exponent and the impact parameter for a null circular unstable geodesic. We see that the increase in $\Lambda$ implies an increasing of the absorption cross--section in both cases ($\omega=-1/2$ and $\omega=-2/3$), then the cosmological constant attenuates the effect of the quintessence, also in both cases the absorption cross--section decreases asymptotically respect to frequency.

\section*{ACKNOWLEDGMENT}

The authors acknowledge the financial support from PROMEP project UAEH--CA--108 and  SNI--CONACYT, M\'exico.

\bibliographystyle{unsrt}

\bibliography{bibliografia}

\end{document}